\documentclass[a4paper,11pt]{article}
\usepackage{jheppub} 

\usepackage{graphicx}
\usepackage{dcolumn}
\usepackage{bm}
\usepackage{mathtools}
\usepackage{amsmath}

\usepackage{xcolor}
\usepackage{caption}
\usepackage{subcaption}
\usepackage{multirow}
\usepackage{booktabs}

\DeclareMathOperator{\pT}{\mathit{p}_{\mathrm{T}}}
 \newcommand{\rev}[1]{{\color{black} #1}}

\title{Jet Rotational Metrics}

\author{Alexis Romero}
\author{and Daniel Whiteson}
\affiliation{Department of Physics and Astronomy, University of California, Irvine, CA, USA 92627}
\emailAdd{alexir2@uci.edu}
\emailAdd{daniel@uci.edu}

\date{\today}

\abstract{
Embedding symmetries in the architectures of deep neural networks can improve classification and network convergence in the context of jet substructure. These results hint at the existence of symmetries in jet energy depositions, such as rotational symmetry, arising from the physical features of the underlying processes. \rev{We introduce new jet observables, Jet Rotational Metrics (JRMs), which provide insights into the substructure of jets by comparing them to jets with perfect discrete rotational symmetry}. We show that JRMs are formidable jet features, achieving good classification scores when used as inputs to deep neural networks. We also show that when used in combination with other jet observables, like N-subjettiness and EFPs, our features increase classification performance. The results suggest that JRMs may capture information not efficiently captured by the other observables, motivating the design of future jet observables for learning the underlying symmetries in the physical processes.
}

\begin{document}
\maketitle
\flushbottom

\section{\label{sec:intro} Introduction}

A prevalent quality of the natural world is the presence of symmetries, from the symmetries in the Standard Model that govern particle interactions at the smallest scales to the isotropy and homogeneity of the universe at the largest scales. The physics at the Large Hadron Collider (LHC) is no exception -- it is governed by the symmetries of the underlying physical processes and the geometry of the detector. 

Deep learning has become a powerful tool for identifying and exploiting symmetries in LHC data. For example, symmetries have been embedded into the architectures of deep neural networks~\cite{bogatskiy2020lorentz,Kasieczka_2019,Gong_2022,Shmakov_2022,Qiu_2023,shimmin2021particle}, generally resulting in better-performing networks with faster convergence. These networks, however, operate on low-level data, such as constituent towers and reconstructed tracks, which tend to be high in volume and dimensionality, making it harder for physicists to interpret results and account for statistical uncertainties. Having compact, high-level features that capture the symmetries in the data could increase the performance of jet classification studies while helping physicists quantify the statistical uncertainties of the data and the models.

A common approach is to treat symmetries in a binary fashion and determine whether data possesses a given symmetry. But in reality, symmetries exist along a spectrum, and data may possess a given symmetry in varying degrees~\cite{DBLP:journals/corr/abs-2002-08791}. Such notions have long been used in chemistry to quantify symmetric properties in molecules~\cite{zabrodsky1992continuous,CSM_water,CSM_chem}, but their applications have less widely been used in areas of particle physics.

\rev{We introduce a new type of jet observables, \textit{Jet Rotational Metrics} (JRMs)\footnote{We call them ``Jerms".}. JRMs are inspired by the discrete, $n$-fold rotational ($C_n$) symmetric form of the Continuous Symmetry Measures~\cite{zabrodsky1992continuous}, which quantify the distance of a given shape from a chosen element of $C_n$-symmetry}. This idea is applied to particle physics by quantifying the similarity $(S)$ between a jet ($J$) and a reference jet ($J_{n}$) with $n$ constituents arranged to have exact $C_{n}$ symmetry:
\begin{equation}
    \mathrm{JRM}_n(J) \propto S(J, J_{n}).
\end{equation}
The construction of $J_{n}$ is discussed in Sec.~\ref{sec:JRMs_rotation}. 

We emphasize that, despite their name, JRMs should not be interpreted as measures of the rotational symmetry of jets. Such measures, if feasible, would need to account for all infinite possible choices of reference jets that are $C_{n}$-symmetric. Instead, JRMs should be interpreted as features of a given $C_{n}$-symmetric element, $J_{n}$, which is carefully designed to resemble the input jet.

\begin{figure}[!ht]
    \centering
    \begin{subfigure}[b]{0.95\textwidth}
        \includegraphics[width=\textwidth]{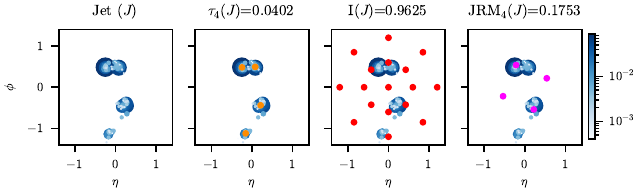}
        \caption{Comparison of the various observables calculated on a jet ($J$).}
        \label{fig:og_jet}
    \end{subfigure}

    \vspace{1cm}
    
    \centering
        \begin{subfigure}[b]{0.95\textwidth}
        \includegraphics[width=\textwidth]{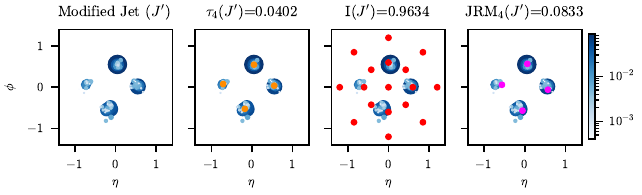}
        \caption{Comparison of the various observables calculated on a jet ($J'$), which has been modified to be more $C_4$-symmetric.}
        \label{fig:modif_jet}
    \end{subfigure}
    \caption{Panel (a) shows a four-prong jet ($J$). The second image in this panel shows the four sub-jet axes (orange dots) used to calculate the N-subjettiness $\tau_4$ measure. The third image shows the isotropic event, $J_{\mathrm{iso}}$ (red dots), used in the calculation of the isotropy measure; $\mathrm{I}(J) \equiv \mathrm{EMD} (J, J_{\mathrm{iso}})$~\cite{cesarotti2020robust}, where $\mathrm{EMD}$ is the Energy Mover's Distance~\cite{Komiske_2019_MSC}. Lastly, the fourth image shows the $C_4$-symmetric event, $J_4$ (pink dots), used in the calculation of the JDM$_4$; JDM$_4 \equiv S(J, J_4)$. The images in panel (b) show similar events but with the jet modified to have a higher degree of $C_4$ symmetry ($J'$) while maintaining the same $\tau_4$ measure. The area and color intensity of the jets' constituents is proportional to their $\pT$.
    }
    \label{fig:feat_comparisons}
\end{figure}

\rev{Comparing jets to reference events is not a new idea. Observables like event isotropy~\cite{cesarotti2020robust} provide similar measures by estimating the amount of work it would take to transform an event into a reference one with isotropic symmetry. Such observables could be helpful when identifying jets that are expected to have high degrees of isotropy, like quark- and gluon-jets. In this paper, we argue that utilizing only one type of symmetry in the reference jets, such as tightly-packed events with an isotropic constituent dispersion, may fail to capture more granular details of the jet substructure -- details that could be useful when identifying jets that are not isotropic but instead display multiple areas rich in harder constituents.

We propose a family of observables whose reference jets are designed to have discrete rotational symmetry, $C_{n}$. Given an ensemble of various $n$s, the observables could form a more complete yet granular picture of the distribution of energy deposition patterns inside jets. An example of the importance of granularity in the reference jets is shown in Fig.~\ref{fig:feat_comparisons}. In this figure, we compare N-subjettiness ($\tau$), a measure of isotropy ($\mathrm{I}$) similar to those introduced in~\cite{cesarotti2020robust}, and JRMs. Panel (a) shows the values of the various observables when calculated on a four-prong jet ($J$). The same jet is modified in panel (b) to have a more symmetric constituent dispersion while maintaining the same $\tau_4$ value. This modification results in a similar isotropy measure as the original jet, while the JRM measure shows a more dramatic difference, decreasing to more than half its original value. In this example, the isotropy measure does not capture the change in constituent dispersion as well as the JRM measure. 
} 

The classification power of JRMs is tested on the benchmark dataset from~\cite{Lu_2022}, which probes their ability to classify jets with multiple sub-jets. We show that JRMs achieve a good classification performance on their own. We also show that when combined with N-subjettiness and EFPs observables, JRMs increase the classifier's performance significantly, surpassing the accuracy of two low-level networks, a PFN and a Transformer. The N-subjettiness and EFP families of observables form complete bases of IRC-safe information and, in principle, a combination of such observables could \rev{learn all relevant jet substructure information}. In practice, however, classifiers operating on these observable families tend to saturate quickly with the number of input features~\cite{Komiske_2018_efp}, meaning they may not efficiently capture all information. Evidence of this is in the boost in performance obtained by supplementing the N-subjettiness and EFPs with JRMs.

The paper is organized as follows: Sec.~\ref{sec:similarity} introduces the similarity measure used to compare the jets. Sec.~\ref{sec:JRMs_rotation} details the procedure \rev{to construct $J_n$ reference jets} and to calculate the JRM observables. Sec.~\ref{sec:Nprong} presents the results of using JRMs as high-level features for jet classification in the benchmark dataset. Concluding remarks and discussion are given in Sec.~\ref{sec:conclusion}.

\section{\label{sec:similarity} Similarity Measure}

JRMs require a similarity measure between two jets. If we consider jets as point clouds, many distance metrics could be used, such as the Chamfer distance~\cite{wu2021densityaware}, the Hausdorff distance~\cite{7053955}, the Earth Mover's distance~\cite{5459199} or its physics-inspired modification, the Energy Mover's distance (EMD)~\cite{Komiske_2019_MSC}. The last measures the work necessary to transform one event into another by solving a system of linear optimal transport equations. Any of these metrics could be used in the calculation of the JRMs, but many are computationally expensive\footnote{We performed a test comparing the results of JRMs using EMD vs. the simple similarity measure used in this paper and found them to be comparable in performance, but the EMD is more computationally expensive. See Appendix~\ref{app:EMD}.}. A useful, cost-effective alternative is to sum the $\pT$-weighted distance between nearest neighbors:
\begin{equation}
    \mathrm{S}(J, J_{n})^{\beta} = \sum \limits_{i \in J} \pT_i \min_{j \in J_{n}} \left( \frac{ \Delta R_{i, j} }{R} \right)^{\beta},
\label{eq:jet_dist}
\end{equation}
where $i$ and $j$ index over the constituents of $J$ and $J_{n}$, respectively. Each constituent $i$ in $J$ queries its nearest neighbor in $J_{n}$. For massless constituents, the pairwise distance $\Delta R_{i, j}$ is their distance in the $(\eta, \phi)$ plane\footnote{Or in the $(y, \phi)$ plane if constituents are not massless.}, which is normalized by the jet radius $R$. The distance between nearest neighbors is weighted by $\pT_i$, the transverse momentum of constituent $i$. The free parameter $\beta$ controls the weight of the angular terms.

Eq.~\ref{eq:jet_dist} resembles the clustering inertia used to measure how well a dataset is clustered by the centroids in $k$-means clustering, which is also used in N-subjettiness. For simplicity, we refer to this measure as the ``distance" between $J$ and $J_{n}$, though this is not a true distance function as the triangle inequality does not hold.

The distance measure defined in Eq.~\ref{eq:jet_dist} is IRC-safe (for $\beta > 0$), but can be generalized to non-IRC-safe versions by varying the exponent of the energy term:
\begin{equation}
    \mathrm{S}(J, J_{n})^{\kappa, \beta} = \sum \limits_{i \in J} \pT_i^{\kappa} \min_{j \in J_{n}} \left( \frac{ \Delta R_{i, j} }{R} \right)^{\beta}.
\label{eq:jet_dist_kb}
\end{equation}
Small $\kappa$ values ($\kappa < 1$) enhance the contributions of softer constituents while large $\kappa$ values ($\kappa > 1$) dampen their contributions. Varying $\kappa$ can provide a sense of the relative dispersion of the softer/harder constituents. Similarly, varying $\beta$ can provide a sense of the relative importance of the geometric terms.

For consistency, we preprocess jets to have net $\pT$ of unity and center them on their $\pT$-weighted mean in the $(\eta, \phi)$ plane. For each constituent $i \in J$:

\begin{align}
    \pT_i &\rightarrow \pT_i / \sum \limits_{j \in J} \pT_j, \\
    \eta_i &\rightarrow  \eta_i - \sum \limits_{j \in J} \pT_j \eta_j / \sum \limits_{j \in J} \pT_j, \\
    \phi_i &\rightarrow  \phi_i - \arctan(\sum \limits_{j \in J} \pT_j \sin(\phi_j) / \sum \limits_{j \in J} \pT_j \cos(\phi_j)) / \sum \limits_{j \in J} \pT_j,
\end{align}

By normalizing the $\pT$, Eq.~\ref{eq:jet_dist} is dimensionless and lies in the range $[0, 1]$. $\mathrm{S}(J, J_{n}) 
= 0$ indicates that the constituents in $J$ are perfectly arranged like those in $J_{n}$ while $\mathrm{S}(J, J_{n}) \approx 1$ indicates a dissimilar arrangement.

\section{\label{sec:JRMs_rotation} 
\rev{Constructing \texorpdfstring{$J_n$}{Jn}}
}

We recall that JRMs measure the similarity between a jet $J$ and a reference jet $J_{n}$ with $C_n$ symmetry. A potential obstacle is that there are infinite possible choices of $J_{n}$, each with perfect $C_n$ symmetry but distinct elements. Ideally, one would choose the reference jet that is closest to $J$, minimizing the distance metric. In practice, an exhaustive search would be computationally prohibitive. To simplify the calculation, we develop a search recipe for the reference jet:
\begin{enumerate}
    \item Consider $n$ points in a circle centered at $(\eta, \phi) = (0, 0)$. The points are located at the angles $2 \pi i / n$, $i=0, \ldots, n-1$. These points represent the $n$ constituents of the reference jet.
    \item Let the points be located at a distance equal to the $\pT$-weighted mean constituent radius ($\Bar{r}$). 
    \item Rotate the points by a common $\theta$ to minimize the distance between $J$ and $J_{n}$, as measured by Eq.~\ref{eq:jet_dist_kb}.
\end{enumerate}
Following the steps specified above, $J_{n}$ is characterized by three parameters: the number of points  (constituents) $n$, the radius $\Bar{r}$, and the rotation angle $\theta \in [0, 2\pi / n)$. With this in mind, JRMs can then be defined as 
\begin{equation}
    \mathrm{JRM}_n^{\kappa, \beta} \coloneqq \min \limits_{\theta} \mathrm{S}(J, J_{n}(\Bar{r}, \theta))^{\kappa, \beta}.
\end{equation}
For clarity, the measure with $\kappa=1$ and $\beta=1$ is simply written as $\mathrm{JRM}_n$. See Appendix~\ref{app:Js_construction} for further discussion about the construction of $J_{n}$. 

\begin{figure}[!ht]
    \centering
    \includegraphics[width=0.9\textwidth]{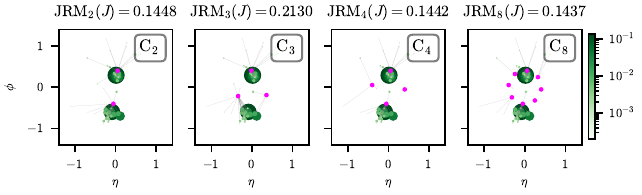}
    \caption{Illustration of the various JRM$_n$ observables calculated on the same two-prong jet ($J$) in green. The area and color intensity of the constituents is proportional to their $\pT$. The pink dots represent the constituents of the $C_n$ symmetric jet ($J_n$), which is compared against $J$. The gray lines depict the nearest neighboring constituents between $J$ and $J_n$. See Section~\ref{sec:Nprong} for details about the jet's topology and generation.}
    \label{fig:JRM2}
\end{figure}

\begin{figure}[!ht]
    \centering
    \includegraphics[width=0.9\textwidth]{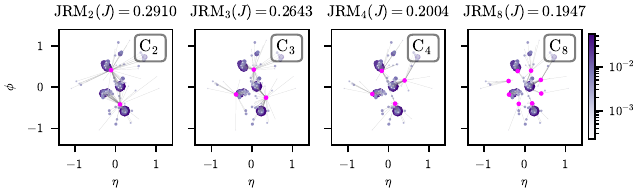}
    \caption{Illustration of the various JRM$_n$ observables calculated on the same eight-prong jet ($J$) in purple. The area and color intensity of the constituents is proportional to their $\pT$. The pink dots represent the constituents of the $C_n$ symmetric jet ($J_n$), which is compared against $J$. The gray lines depict the nearest neighboring constituents between $J$ and $J_n$. See Section~\ref{sec:Nprong} for details about the jet's topology and generation.}
    \label{fig:JRM8}
\end{figure}

An illustration of JRMs with $n=2,3,4,8$ for a two-prong jet is shown in Fig.~\ref{fig:JRM2}. Interestingly, $\mathrm{JRM}_3 > \mathrm{JRM}_2$, which makes sense considering how the constituents of the jet are mainly concentrated in two areas, and it would be hard to compare them to a reference jet with three equidistant constituents. The same observables are illustrated in Fig.~\ref{fig:JRM8} for an eight-prong jet. The constituents in the jet are distributed more uniformly, which makes the JRM measure decrease with $n$; \rev{$\mathrm{JRM}_{n} > \mathrm{JRM}_{n+1}$. Like the N-subjettiness variables, it is useful to analyze not a single JRM but an ensemble and their relative values}. An analysis of JRM ratios is shown in Appendix~\ref{app:ratios}.

In Fig.~\ref{fig:JRM2}, we note how similar JRM$_2$, JRM$_4$, and JRM$_8$ are for the two-prong jet. Because of the nearest neighbor operation, the majority of the constituents are assigned to only two reference constituents in the JRM$_4$ and JRM$_8$ observables. While this may seem like a weakness, we find that attempts to penalize the measure for an unbalanced constituent assignment result in slightly lower performances by the classifiers. In addition, \rev{in our studies, we corroborate our results by also calculating JRM observables using EMD as the similarity measure. EMD results in different values of JRM$_2$, JRM$_4$, and JRM$_8$, which is expected. However, when used as input to a neural network, the performance is similar to the JRM observables using the simpler similarity measure from Eq.~\ref{eq:jet_dist_kb} is comparable to those using the more expensive EMD. See Appendix~\ref{app:EMD} for further discussion.}

The results suggest that JRMs may help a neural network learn details of the energy deposition patterns in jets, even when the nearest neighbor operation assigns most constituents to a few reference constituents. A network may find value in which $n$s result in similar or different JRM measures. An example of this could be how the observables of the form JRM$_{2n}$ have similar values in Fig.~\ref{fig:JRM2}, which suggests that the constituents in the jet are arranged in two quasi-symmetric clusters.

\section{Multi-Prong Jet Classification}
\label{sec:Nprong}

\begin{figure}[!ht]
    \centering
    \includegraphics[width=0.55\textwidth]{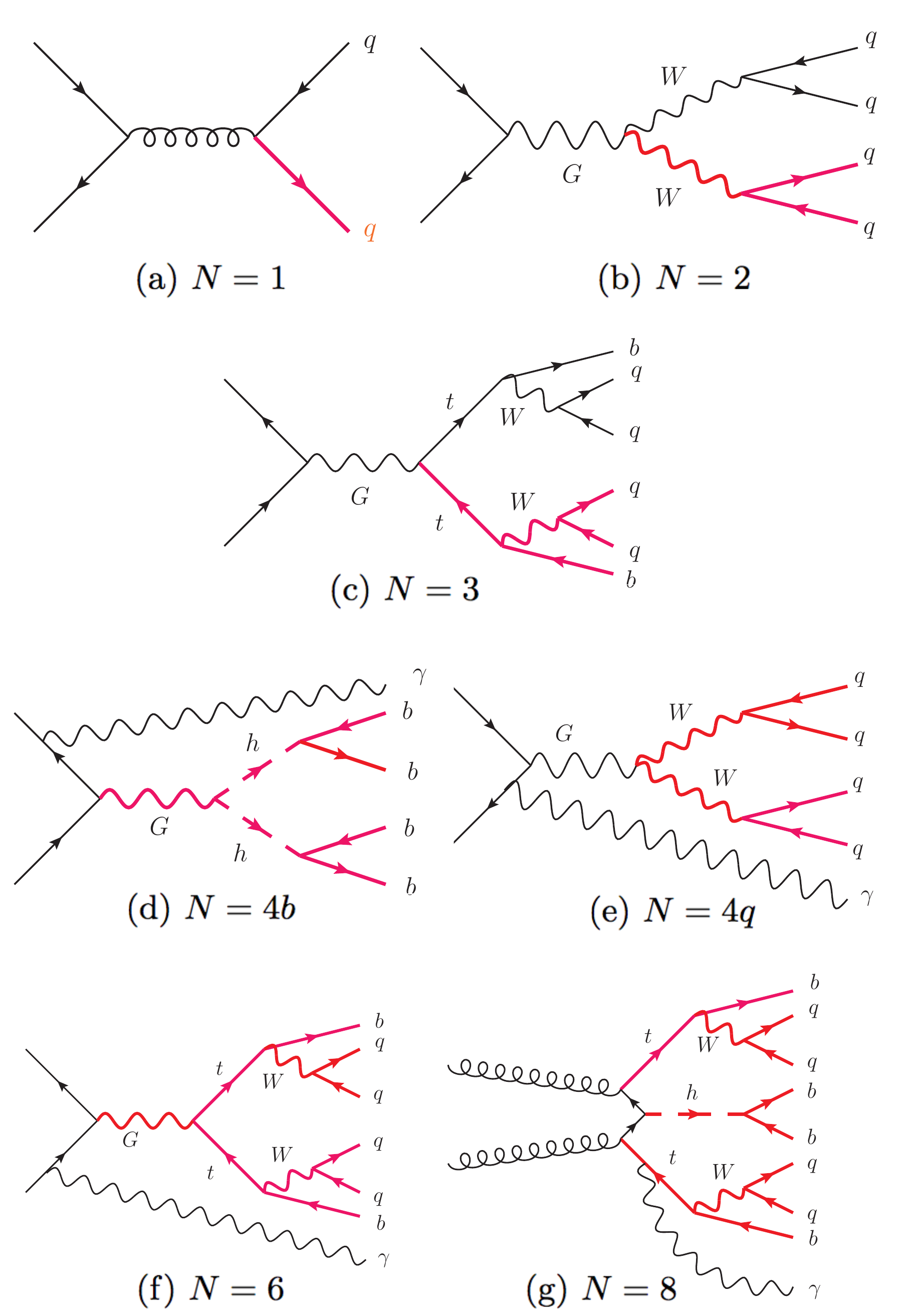}
    \caption{Feynman diagrams for the multi-prong dataset processes that generate jets with $N=1,2,3,4b, 4q, 6,8$ hard sub-jets. The lines in red indicate the components, which are required to be truth-matched to the jet.}
    \label{fig:diagrams}
\end{figure}

In this section, we evaluate the performance of JRMs in the classification of jets with multiple sub-jets using the benchmark dataset from Ref.~\cite{Lu_2022}. Jets are generated with $N=1, 2, 3, 4, 6, 8$ hard sub-jets. The $N=4$ class is subdivided into $N=4q$ and $N=4b$, for a total of seven classes. The Feynman diagrams of the processes used to generate the jet samples are shown in Fig.~\ref{fig:diagrams}. 

The dataset consists of 108,359 simulated jets, which we divide into training, validation, and test sets with proportions $80:10:10$, respectively.

\begin{figure}
    \centering
    \includegraphics[width=\textwidth]{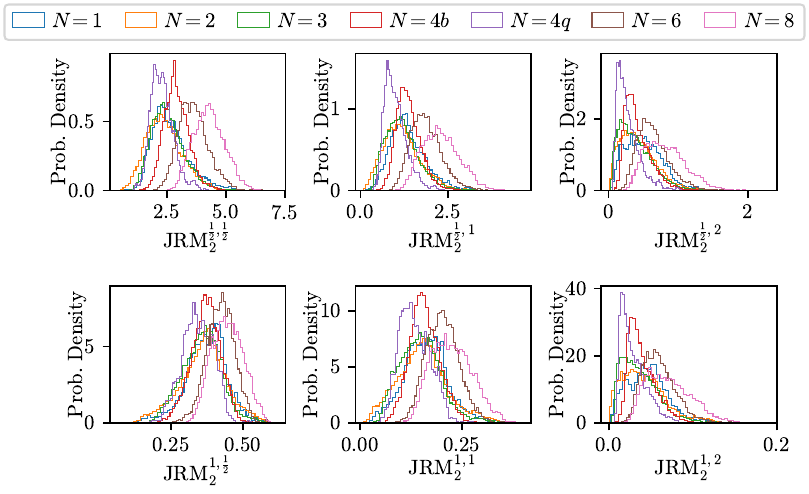}
    \caption{Distributions of the JRM$_2^{\kappa, \beta}$ observables with the specified $\kappa$ and $\beta$ parameters.}
    \label{fig:C2_dists}
\end{figure}

We calculate a total of 30 $\mathrm{JRM}$ observables with the following parameter combinations\footnote{We utilized the same $n$ values as the number of sub-jets, which perform well for the classification task, though this may not always be the case. The optimal JRM feature for identifying jets with $N$ sub-jets may have $n \neq N$.}: $n=2, 3, 4, 6, 8$, $\kappa = \frac{1}{2}, 1$, and $\beta = \frac{1}{2}, 1, 2$. As an example of these observables, we show the distributions of the JRM$_2$ features in Fig.~\ref{fig:C2_dists}. It is apparent that the JRM$_2$ observables provide useful insights into the substructure of the jets. For example, the $N=2$ and $N=4q$ classes have the lowest mean JRM$_2$ values. Conversely, the $N=6$ and $N=8$ classes have the highest mean JRM$_2$ values, which suggests that the constituents in these classes are less likely to be collimated in such a way that they are concentrated in two symmetric areas. 

\begin{figure}
    \centering
    \includegraphics[width=\textwidth]{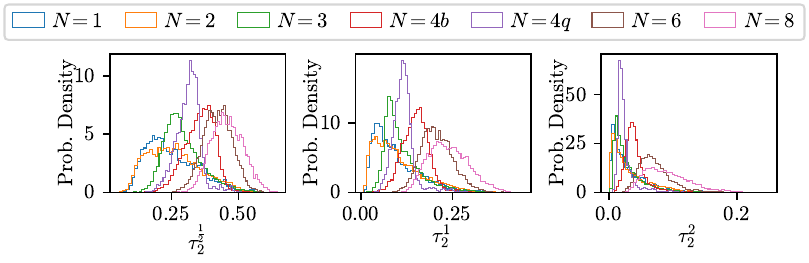}
    \caption{Distributions of the N-subjettiness features with N=2 ($\tau_2^{\beta}$) observables with the specified $\beta$ parameters.}
    \label{fig:tau2_dists}
\end{figure}

To highlight the differences between JRM$_2$ and the 2-axis N-subjettiness features ($\tau_2^{\beta}$), we show the latter in Fig.~\ref{fig:tau2_dists}. The classes with the lowest $\tau_2$ values are $N=1$ and $N=2$, while the $N=4q$ class falls in the middle of all classes. Similarly to JRM$_2$, the $N=6$ and $N=8$ classes have the largest $\tau_2$ values.

The rest of the distributions are shown in Appendix~\ref{app:distributions}. In our experiments, JRMs nearly saturate for $n=8$. This is seen in the distributions in Appendix~\ref{app:distributions}, where the JRM$_6$ measures resemble JRM$_8$. In addition, including JRM$_n$ observables with $n > 8$ does not increase the classifiers' performance.

To assess the performance of these observables and the uniqueness of the information they capture, we compare the performance of dense networks operating on JRMs to those operating on the N-subjettiness and EFP observables from Ref.~\cite{Lu_2022}\footnote{The studies in~\cite{Lu_2022} include jet mass and constituent multiplicity, but we omit those variables here and focus on comparing the predictive power of JRMs against N-subjettiness and EFP variables.}. The set of N-subjettiness variables consists of 135 observables calculated along the $k_{\mathrm{T}}$ axis, with sub-jet axis parameter $\mathrm{N}=1, \ldots, 45$ and angular weighing parameter $\beta = \frac{1}{2}, 1, 2$.  The EFP set consists of 162 observables representing prime multi-graphs with five or fewer edges, energy weighing parameter $\kappa=1$ and angular weighing parameter $\beta = \frac{1}{2}, 1, 2$. We consider both sets to be extensive, as including more N-subjettiness or EFP observables is not likely to increase the performance of the networks, and if so, marginally\footnote{As a cross-check, we trained networks on half as many N-subjettiness and EFPs with $d=5$ edges and found almost no drop in the classification score, which suggests that the considered set may be saturating the classifiers.}.

For consistency and to test the boost in performance obtained by including JRMs, all dense networks share the same architecture of six hidden layers of size (800-800-800-800-800-64) and \textsc{ReLu}~\cite{inproceedings} activation function. Dropout with a rate of 0.3 and batch normalization are applied sequentially after every hidden layer. The output layer has dimension seven and a softmax activation function\footnote{The network hyperparameters were chosen by a grid search.}. 

To gain a better understanding of the nature of the information used by the models, we distinguish between IRC-safe and IRC-unsafe observables. Table~\ref{tab:nprong_acc_safe} shows the results of the networks operating on IRC-safe observables. The DNN trained on IRC-safe JRMs achieves an overall accuracy of $85.49 \pm 0.35\%$. The DNN trained on the larger set 297 of N-subjettiness and EFPs exceeds that, with an overall accuracy of $88.94 \pm 0.37\%$. The highest accuracy of $90.72 \pm 0.23\%$ is achieved by combining all observables, which suggests that the JRMs could be helping the network learn information more efficiently. This information is highly valuable for the classification process since it not only increases the overall accuracy by more than $1.5\%$ but also the accuracy of each class. See Appendix~\ref{app:per_class_acc} for the classification accuracies per class.   

Table~\ref{tab:nprong_acc_all} shows the results of the models whose input is IRC-agnostic. The PFN and Transformer models from~\cite{Lu_2022} are used as the benchmark low-level networks to compare against the high-level features. The DNN trained on the full set of 30 JRMs does very well, achieving an overall accuracy of $89.30 \pm 0.26\%$ and outperforming the overall accuracy of the PFN~\cite{Komiske_2019_EFN} ($89.19 \pm 0.23\%$), which operates on low-level information -- constituent momenta. More significant is the boost in performance obtained when combining all the JRMs with the N-subjettiness and EFPs, which together achieve the top overall accuracy of $91.52 \pm 0.26\%$ and outperform the Transformer ($91.27 \pm 0.31\%$) operating on low-level information. By including the IRC-unsafe JRM observables, the gap between the Transformer and dense networks is bridged. This suggests that deep, low-level networks trained on the full event information may make use of soft and collinear radiation, which is not well-defined in the perturbative QCD regime. Moreover, these results provide insights into $\textit{how}$ this information is used. \rev{Low-level networks may be learning not only the number of clusters in which the jets' constituents are collimated but also the relative distribution between clusters; for example, whether the clusters are dispersed equidistantly like the reference jets.} We note that the contributions of the IRC-unsafe features are small when compared to the classification power of the IRC-safe ones, which is expected, but they cannot be discarded when accounting for statistical uncertainty of the low-level networks due to deviations in the particle simulations~\cite{Gras_2017}.

\begin{table}[!ht]
    \centering
     \caption{Mean 10-fold prediction accuracy and statistical uncertainty of the DNNs operating on the specified IRC-safe observables.}
    \label{tab:nprong_acc_safe}
    \begin{tabular}{llrc}
        \hline \hline
             Network &
             Input (IRC-safe) &
             Input Dim. &
             Overall Acc. (\%) \\
        \hline
            DNN &
            JRM & 
            15 &
            85.49 $\pm$ 0.35 
            \\ 
            DNN &
            $\mathrm{N}$-subs, EFPs &
            297 &
            88.94 $\pm$ 0.37
            \\
            DNN &
            $\mathrm{N}$-subs, EFPs, JRMs &
            312 &
            \textbf{90.72} $\pm$ 0.23
            \\
            \hline\hline
    \end{tabular}
\end{table}

\begin{table*}[ht]
    \centering
    \caption{Mean 10-fold prediction accuracy and statistical uncertainty of the various networks. The input to the PFN and Transformer is the constituents' three-momentum, which is zero-padded to have a uniform length of 230. See~\cite{Lu_2022} for further details about the PFN and Transformer networks.}
    \label{tab:nprong_acc_all}
    \begin{tabular}{llrc}
        \hline \hline
             Network &
             Input &
             Input Dim. &
             Overall Acc. (\%) \\
        \hline
            DNN &
            JRMs &
            30 &
            89.30 $\pm$ 0.26 
            \\
            DNN &
            $\mathrm{N}$-subs, EFPs, JRMs &
            327 &
            \textbf{91.52} $\pm$ 0.26 
            \\
            PFN &
            Constituents &
            (230, 3) &
            89.19 $\pm$ 0.23
            \\
            Transformer &
            Constituents &
            (230, 3) &
            91.27 $\pm$ 0.31
            \\
        \hline\hline
    \end{tabular}
\end{table*}

\subsection{Feature Selection}

We continue our exploration of JRMs for multi-prong classification by identifying their relevant importance in the classification task. We employ a simple feature selection method based on $L_1$~\cite{lasso} regularization, also known as LASSO regularization. This type of regularization is commonly used in linear regression for feature selection. The core idea behind it is to penalize the loss function of the regression by adding a term proportional to the absolute values of the weights of the linear terms, which encourages irrelevant weights to be zeroed out. Here, we penalize the loss function of the dense network by adding a term proportional to the weights of the input features. 

We perform feature selection on the 312 IRC-safe observables (N-subs, EFPs, and JRMs). The input layer of the dense network has dimension 312, and it is fully connected to the first hidden layer of size 800, meaning that feature $i$ has weights $\bm{w}_i = [w_1, \ldots, w_{800}]^{T}$. We implement a learnable gate parameter $g_i$ that multiplies the weight vector of feature $i$; $\bm{w}_i \rightarrow g_i \bm{w}_i$. The LASSO regularization is applied to the gate parameters such that if $g_i$ is zeroed out, so is the entire weight vector of feature $i$. The loss function ($L$) of the dense network can then be written as
\begin{equation}
    L = -\mathrm{log} f (Y, Y_{\mathrm{pred}}) + \lambda \sum \limits_{i=1}^{312} \mid g_i \mid.
\end{equation}
The first term corresponds to the negative log-likelihood of the true ($Y$) and predicted ($Y_{\mathrm{pred}}$) labels. The second term is the LASSO penalization on the gate parameters with regularization strength parameter $\lambda$. We set $\lambda=5$, and to further reduce the size of the subset, we keep only the observables with a gate parameter $\mid g_i \mid > = 0.01$. 

The feature selection method has identified 29 IRC-safe observables, which are shown in Table~\ref{tab:lasso_features_IRCsafe}. A DNN trained on these observables reaches an accuracy of $89.97 \pm 0.28\%$, well approximating the accuracy of the full set and slightly surpassing the accuracy of the PFN. These results are impressive given the limited number of input features. We note that despite the successful results of the LASSO-inspired method, it does not guarantee that the subset of features selected is the ideal one for the classification task. Finding such a subset would require a more in-depth analysis of the features and their correlations.

\begin{table}[!ht]
    \centering
    \caption{IRC-safe observables chosen by the LASSO-inspired feature selection method, which achieve an overall accuracy of $89.97 \pm 0.28\%$ when used as input to a DNN. The selected N-subjettiness observables, $\tau^{\beta}_\mathrm{N}$, are shown in the leftmost column. The selected JRM observables are shown in the middle column. The selected EFP observables, EFP$^{\beta}$(n,d,k), are shown in the rightmost column and are accompanied by their unique identifiers indicating the number of nodes (n), edges (d), and index (k); see~\cite{Komiske_2018_efp}. The observables are listed in no particular order.}
    \label{tab:lasso_features_IRCsafe}
    \begin{tabular}{lll}
        \hline \hline
        N-subjettiness & JRMs & EFPs \\
        \hline
        $\tau^{2}_{2}$        & JRM$\rev{_2}^{1, 1}$      & EFP$^{2}(2, 2, 0)$     \\
        $\tau^{0.5}_{3}$      & JRM$\rev{_3}^{1, 0.5}$    & EFP$^{2}(2, 5, 0)$     \\
        $\tau^{1}_{3}$        & JRM$\rev{_3}^{1, 2}$      & EFP$^{0.5}(3, 3, 1)$   \\
        $\tau^{1}_{4}$        & JRM$\rev{_3}^{1, 1}$      & EFP$^{2}(3, 4, 3)$     \\
        $\tau^{1}_{5}$        & JRM$\rev{_4}^{1, 0.5}$    & EFP$^{2}(4, 5, 0)$     \\
        $\tau^{1}_{7}$        & JRM$\rev{_4}^{1, 1}$      & EFP$^{2}(4, 5, 3)$     \\
        $\tau^{0.5}_{16}$     & JRM$\rev{_6}^{1, 0.5}$    & \\
        $\tau^{0.5}_{22}$     & JRM$\rev{_6}^{1, 2}$      & \\
        $\tau^{0.5}_{39}$     & JRM$\rev{_8}^{1, 1}$      & \\
        $\tau^{0.5}_{39}$     &                       & \\
        $\tau^{0.5}_{45}$     &                       & \\            
        \hline \hline
    \end{tabular}
\end{table}

The choice of features provides insights into the classification strategies of the dense networks. The family of observables with the most selected features is N-subjettiness, and most of these variables have N$<8$, which makes sense given the pronginess of the jets in the dataset\footnote{This result is consistent with~\cite{Lu_2022}, where N-subjettiness variables with $N<8$ were also found to be most relevant for the classification task.}. The next family of observables is JRMs. \rev{For every $n$ considered, at least one JRM$_n$ is chosen. These results suggest that not only JRMs are useful to the network, but also their combination}. The feature selection method also chooses a fair amount of EFP variables with 2, 3, and 4 nodes. Some of these observables correlate to other well-known jet observables, such as EFP(2, 1, 0) and the jet mass~\cite{Komiske_2018_efp}.

\section{\label{sec:conclusion} Conclusions}

\rev{In this paper, we have introduced Jet Rotational Metrics, JRMs, a new type of jet observables. The core idea behind JRMs is that the energy deposition patterns in jets are highly dependent on the underlying processes, which may dictate measurable features such as the approximate number of clusters or sub-jets in the jets. Observables like N-subjettiness or EFPs do a good job gauging the number of sub-jets~\cite{Lu_2022}, but they may not be as efficient in capturing \textit{how} these sub-jets are roughly dispersed. For example, some topologies may result in jets with sub-jets that are mostly collimated, while others may result in sub-jets that are more spread apart. A combination of N-subjettiness or EFP observables is likely to capture some of this information, but in practice, classifiers trained on these families of observables tend to saturate quickly with the number of input features~\cite{Komiske_2018_efp}. JRMs aim to efficiently capture this information by comparing jets to reference jets with an equidistant constituent dispersion.

Inspired by the $C_n$-symmetric form of the Continuous Symmetry Measures~\cite{zabrodsky1992continuous}, JRMs measure the similarity between a jet ($J$) and a reference jet ($J_{n}$) with $n$ constituents arranged to have exact $C_{n}$ symmetry. Many similarity measures are considered, but in this paper we opt for a cheap, simple measure that sums the distance from every constituent in $J$ to its nearest neighbor in $J_{n}$. We, however, note that other similarity measures, such as the Earth Mover's Distance, could be used in the calculation of JRMs.}

The JRM observables are tested on a benchmark dataset for classifying jets with multiple sub-jets. More specifically, jets with $N=1, 2, 3, 4, 6, 8$ sub-jets. We find that JRMs are very useful discriminants, achieving formidable classification accuracies when used as input to dense neural networks. Moreover, when combined with more traditional observables, like N-subjettiness and EFP variables, JRMs increase the dense neural network's accuracy, outperforming two benchmark low-level networks. From these results we draw two conclusions: First, JRMs help the network learn information more efficiently than simply training on the traditional features. Second, this information is {\it highly relevant} for the classification task, bridging the gap between high- and low-level networks and shedding light on the learning strategies used by these networks. 

Recent advances in machine learning for jet classification have focused on deep, low-level networks. These networks learn functions directly from the jet constituents, so it is no surprise that they currently provide state-of-the-art results in many  applications~\cite{qu2022particle,Collado:2020ehf,Collado:2020fwm,Witkowski:2023htt, Faucett:2022zie}. However, the high performance of these networks comes with the downside of lower interpretability as it is not clear which specific functions of the complex, low-level input they have learned. 

High-level observables, like JRMs, make it more feasible for physicists to understand the nature of the information learned by the networks. These observables are often used as inputs to dense neural networks, which are easier to interpret since we know the functional forms of their inputs. An example of this is observables that are IRC-safe, an important property to test the substructure of an event while being insensitive to soft and collinear radiation. Although N-subjettiness and EFPs form bases of IRC-safe information, capturing all such relevant information in a small set of observables can be difficult. We show that IRC-safe JRMs are good complements to these more traditional observables and that, together, they have the potential to approximate the performance of low-level networks. 
 
Lastly, we acknowledge that there could be many natural extensions to JRMs, either by utilizing different forms of $C_n$-symmetric elements \rev{in the reference jets} or by employing different distance metrics. We leave further exploration of these choices for future work.

\section*{Acknowledgments} DW is funded by the DOE Office of Science.  The authors thank Chase Shimmin, Adam Rennie, Ben Nachman, and Joakim Olsson for useful comments.

\appendix
\section{\label{app:Js_construction} Further Discussion on the Choice of \texorpdfstring{$J_n$}{Jn}}

We mention in Sec.~\ref{sec:JRMs_rotation} how, ideally, JRMs would measure the distance between a jet $J$ and its closest $C_n$-symmetric element. In practice, however, finding this element would be computationally prohibitive due to the infinite possible ways to construct $C_n$-symmetric elements, all with different numbers of constituents, radii, and orientations. To simplify this task, we developed a recipe to construct the reference jet by (1) considering $n$ equidistant points centered at $(\eta, \phi) = (0, 0)$, (2) placing them at a radius equal to the $\pT$-weighted constituent mean radius, and (3) rotating points to minimize their distance to $J$. The final arrangement of the points represents the constituents in the reference jet, which we call $J_{n}$.

In this appendix, we test whether the aforementioned recipe for $J_{n}$ results in good discriminants or if different choices yield features with better classification performances. We test step (2) by comparing the standard JRMs to similar features where we set the radius equal to the jet radius. The results are shown in the second to last row of Table~\ref{tab:nprong_appx}. These new features achieve an overall accuracy of $87.83 \pm 0.47\%$, which is roughly $1.5\%$ less than that of the standard JRMs. 

We also test step (3) by comparing the standard JRMs to similar features where the \rev{constituents are not rotated by a common angle} $\theta$ to minimize the similarity measure. The results are shown in the last row of Table~\ref{tab:nprong_appx}. These new features achieve an overall accuracy of $87.81 \pm 0.23\%$, which is also roughly $1.5\%$ less than that of the standard JRMs. 

The results suggest that the recipe used to construct $J_{n}$ results in good jet substructure discriminants. We note, however, that while this recipe greatly simplifies the search for $C_n$-symmetric elements, it is not computationally trivial since it requires the calculation of the mean constituent radius and angular optimization. We show that some information is lost when these steps are omitted, resulting in lower accuracies. Still, the simplified features achieve good classification performances.

\begin{table*}[ht]
    \centering
     \caption{Mean 10-fold prediction accuracy and statistical uncertainty of DNNs operating on the specified input features.}
    \label{tab:nprong_appx}
    \begin{tabular}{l|c}
        \hline \hline
            {Input} &
            {Overall Acc. (\%)}\\
            \hline
            Standard JRMs &
            \textbf{89.30} $\pm$ 0.26
            \\
            \rev{JRMs modified so the constit. lie at a dist. equal to the jet radius}&  
            {87.83 $\pm$ 0.47} 
            \\

            JRMs without the $\theta$ optimization &
            {87.81 $\pm$ 0.23} 
            \\
            \hline
    \end{tabular}
\end{table*}

\section{\label{app:EMD}Further Discussion on the Choice of Similarity Measure}

In this paper, we have opted to define the similarity between two jets as the sum of the $\pT$-weighted distance between nearest constituents. This simple measure is chosen because of its simplicity. Jets can have hundreds of constituents, and thus, having a cheap and highly interpretable way of comparing two or more jets can be critical. In addition, this simple measure is comparable to the clustering inertia used in the calculation of N-subjettiness, which makes JRM good supplemental features for high-level analyses using N-subjettiness variables. 

Despite its efficiency, the simple measure is not a true distance metric as it does not necessarily satisfy the triangle inequality. More rigorous distance metrics have been used to compare jets, such as the Energy-Mover's distance (EMD)~\cite{Komiske_2019_MSC}, which is a particle-physics application of the Earth-Mover's distance~\cite{5459199}. The EMD quantifies the work necessary to transform one event into another, making it a good candidate for the choice of similarity measure used in JRMs. 

Table~\ref{tab:EMD_1} shows the overall accuracy of DNNs trained on JRMs, which either are calculated with the choice of the simple similarity measure or with EMD. The results show that the overall accuracy is slightly higher when the JRMs are calculated with the simple similarity measure, which is a little surprising since the EMD is a more rigorous metric. Also shown in the table is the overall accuracy of a DNN trained on both sets of JRMs. The results show a small but non-negligible increase in accuracy, indicating that there may be unique information in the two sets. These results are intriguing, and they invite an exploration of the mutually exclusive information between jet classes and how to quantify it using various similarity measures, which is reserved for future work.

\begin{table*}[ht]
    \centering
    \caption{Mean 10-fold prediction accuracy and statistical uncertainty of the various networks.}
    \label{tab:EMD_1}
    \begin{tabular}{llrc}
        \hline \hline
             Network &
             Input &
             Input Dim. &
             Overall Acc. (\%) \\
        \hline
            DNN &
            JRMs (EMD) &
            30 &
            88.43 $\pm$ 0.33 
            \\
            DNN &
            JRMs (simple) &
            30 &
            89.30 $\pm$ 0.26 
            \\
            DNN &
            JRMs (EMD) and JRMs (simple) &
            60 &
            \textbf{90.91} $\pm$ 0.29 
            \\
        \hline\hline
    \end{tabular}
\end{table*}

\section{\label{app:per_class_acc}Multi-Prong Jet Classification: Per-Class Accuracy}

In this appendix, we show the results of the various networks on each class of the multi-prong classification dataset. The results of the dense networks operating on IRC-safe observables are shown in Table~\ref{tab:nprong_acc_safe_per_class}. The results of the networks operating on IRC-agnostic input are shown in Table~\ref{tab:nprong_acc_all_per_class}.

\begin{table}[!ht]
    \centering
     \caption{Mean 10-fold class prediction accuracy and statistical uncertainty of the DNNs operating on the specified IRC-safe observables. See Table~\ref{tab:nprong_acc_safe} for the overall accuracy values.}
    \label{tab:nprong_acc_safe_per_class}
        \begin{tabular}{l@{\hskip 0.2cm}l@{\hskip 0.05cm}|@{\hskip 0.05cm}c@{\hskip 0.05cm}|@{\hskip 0.05cm}c@{\hskip 0.05cm}|@{\hskip 0.05cm}c@{\hskip 0.05cm}|@{\hskip 0.05cm}c@{\hskip 0.05cm}|@{\hskip 0.05cm}c@{\hskip 0.05cm}|@{\hskip 0.05cm}c@{\hskip 0.05cm}|@{\hskip 0.05cm}c}
        \hline \hline
             \multirow{2}{*}{Net.} &
             \multirow{2}{*}{\shortstack[l]{Input}} &
             \multicolumn{7}{c}{Acc. per Class (\%)} \\
        \cline{3-9} 
                &  &  $N=1$ & $N=2$ & $N=3$ & $N=4b$ & $N=4q$ & $N=6$ & $N=8$ \\ 
        \hline \hline
            DNN &
            JRMs & 
            92.3$\pm$0.8 & 
            79.3$\pm$1.3 & 
            78.6$\pm$1.0 & 
            76.6$\pm$1.1 & 
            97.1$\pm$0.5 &
            91.1$\pm$0.6 & 
            83.2$\pm$1.0 \\ 
            \hline
            \multirow{3}{*}{DNN} &
            \multirow{3}{*}{\shortstack[l]{$\mathrm{N}$-subs,\\EFPs}} &
            \multirow{3}{*}{93.0$\pm$0.6}&
            \multirow{3}{*}{83.0$\pm$0.9}&
            \multirow{3}{*}{82.9$\pm$1.3}&
            \multirow{3}{*}{83.7$\pm$0.5}&
            \multirow{3}{*}{98.0$\pm$0.3}&
            \multirow{3}{*}{92.0$\pm$0.7}&
            \multirow{3}{*}{90.1$\pm$1.2}\\
             & & & & & & & & \\
             & & & & & & & & \\
            \hline
            \multirow{4}{*}{DNN} &
            \multirow{4}{*}{\shortstack[l]{$\mathrm{N}$-subs,\\EFPs,\\JRMs}} &
            \multirow{4}{*}{\textbf{94.6}$\pm$0.5}&
            \multirow{4}{*}{\textbf{85.5}$\pm$1.1}&
            \multirow{4}{*}{\textbf{85.2}$\pm$0.8}&
            \multirow{4}{*}{\textbf{85.1}$\pm$1.3}&
            \multirow{4}{*}{\textbf{98.5}$\pm$0.2}&
            \multirow{4}{*}{\textbf{94.7}$\pm$0.6}&
            \multirow{4}{*}{\textbf{91.4}$\pm$0.8}\\
            & & & & & & & & \\
            & & & & & & & & \\
            & & & & & & & & \\
            \hline\hline
    \end{tabular}
    \clearpage
\end{table}

\begin{table}[!ht]
    \centering
    \caption{Mean 10-fold class prediction accuracy and statistical uncertainty of the various networks. The input to the PFN and Transformer is the constituents' three-momentum, which is zero-padded to have a uniform length of 230. See~\cite{Lu_2022} for further details about the PFN and Transformer networks. The full set of 30 JRMs is used as input to the DNN. See Table~\ref{tab:nprong_acc_all} for the overall accuracy values.}
    \label{tab:nprong_acc_all_per_class}
        \begin{tabular}{l@{\hskip 0.2cm}l@{\hskip 0.05cm}|@{\hskip 0.05cm}c@{\hskip 0.05cm}|@{\hskip 0.05cm}c@{\hskip 0.05cm}|@{\hskip 0.05cm}c@{\hskip 0.05cm}|@{\hskip 0.05cm}c@{\hskip 0.05cm}|@{\hskip 0.05cm}c@{\hskip 0.05cm}|@{\hskip 0.05cm}c@{\hskip 0.05cm}|@{\hskip 0.05cm}c}
        \hline \hline
             \multirow{2}{*}{Net.} &
             \multirow{2}{*}{\shortstack[l]{Input}} &
             \multicolumn{7}{c}{Acc. per Class (\%)} \\
        \cline{3-9} 
                &  &  $N=1$ & $N=2$ & $N=3$ & $N=4b$ & $N=4q$ & $N=6$ & $N=8$ \\ 
        \hline \hline
            DNN &
            JRMs &
            96.0$\pm$0.6& 
            86.6$\pm$1.1& 
            84.2$\pm$0.5& 
            80.1$\pm$0.9& 
            98.2$\pm$0.3& 
            93.4$\pm$0.7& 
            88.4$\pm$1.0\\
            \hline
            \multirow{4}{*}{DNN} &
            \multirow{4}{*}{\shortstack[l]{$\mathrm{N}$-subs,\\EFPs,\\JRMs}} &
            \multirow{4}{*}{95.7$\pm$0.8} & 
            \multirow{4}{*}{87.2$\pm$1.3} & 
            \multirow{4}{*}{86.4$\pm$0.8} & 
            \multirow{4}{*}{\textbf{86.2}$\pm$1.0} & 
            \multirow{4}{*}{\textbf{98.6}$\pm$0.3} & 
            \multirow{4}{*}{\textbf{94.9}$\pm$0.6} & 
            \multirow{4}{*}{\textbf{91.5}$\pm$0.6} \\
            & & & & & & & & \\
            & & & & & & & & \\
            & & & & & & & & \\
            \hline
            PFN &
            Consts. &
            \textbf{96.4}$\pm$0.6&
            85.2$\pm$1.3&
            82.9$\pm$1.1&
            78.8$\pm$0.5&
            98.0$\pm$0.3&
            93.4$\pm$0.9&
            89.7$\pm$0.9\\
            \hline
            Trans. &
            Consts. &
            96.0$\pm$0.5&
            \textbf{88.1}$\pm$1.2&
            \textbf{86.7}$\pm$1.1&
            84.6$\pm$1.2&
            98.5$\pm$0.3&
            94.0$\pm$1.1&
            91.1$\pm$1.2\\
        \hline\hline
    \end{tabular}
\end{table}

\section{\label{app:ratios} JRM Ratios}

In this paper, we use the high-level observables as input to dense networks, but a common approach is to utilize their ratios in statistical-cut-based analyses. For example, ratios of $\tau_\mathrm{N} / \tau_{\mathrm{N}-1}$ N-subjettiness observables have been used to classify jets with different topologies~\cite{Thaler_2012_ratios,Napoletano_2018}. A jet that is very ``N-subjetty" will show a relatively large difference between the $\tau_\mathrm{N}$ and $\tau_{N-1}$ variables, which will be reflected in their ratio. 

\begin{figure}[!ht]
    \centering
    \includegraphics[width=0.8\textwidth]{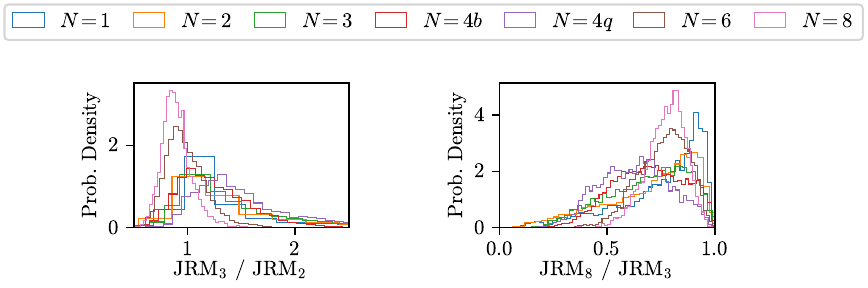}
    \caption{Distributions of the selected JRM ratios.}
    \label{fig:ratios}
\end{figure}

Because of the nature of JRMs, we do not expect ratios of JRM$_n$/JRM$_{n-1}$ to vary greatly for large $n$. Useful ratios could be JRM$_3$/JRM$_2$, and between JRM$_3$ and a larger $n$ that captures features of the isotropy of the jets, such as JRM$_8$/JRM$_3$. Distributions of such ratios are shown in Fig.~\ref{fig:ratios}. We do not perform an analysis using statistical cuts on the ratios, but their distributions show a good degree of class separation.

\section{\label{app:distributions}JRM Distributions}

\begin{figure}[!ht]
    \centering
    \includegraphics[width=\textwidth]{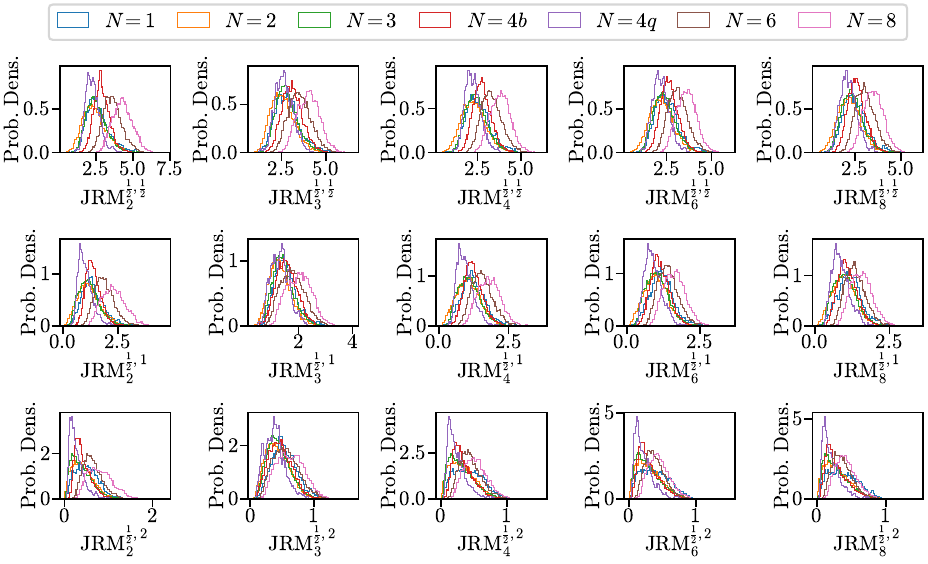}
    \caption{Distributions of the JRM$_n^{\kappa, \beta}$ features with $\kappa=\frac{1}{2}$, and the specified $n$ and $\beta$ parameters. The rest of the distributions are shown in Fig.~\ref{fig:Cn_dists_2}.}
    \label{fig:Cn_dists_1}

    \centering
    \includegraphics[width=\textwidth]{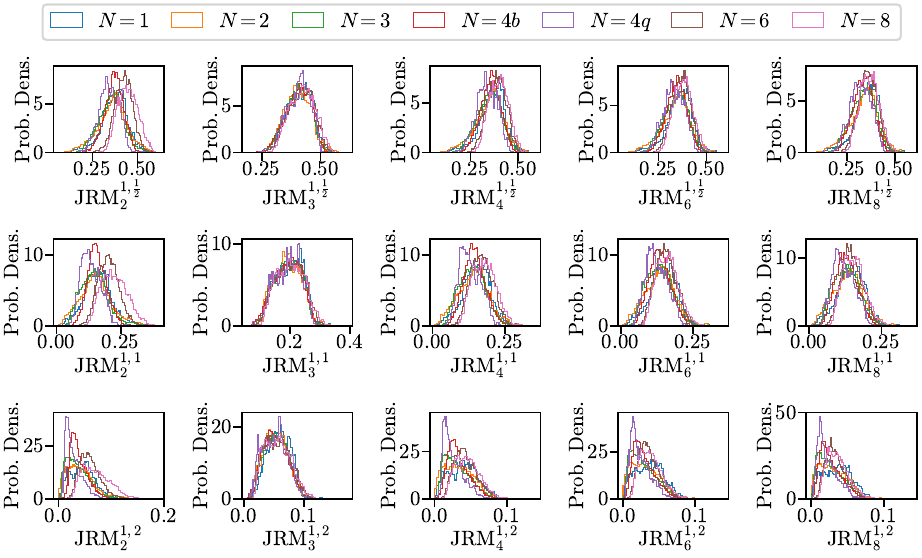}
    \caption{Distributions of the JRM$_n^{\kappa, \beta}$ features with $\kappa=1$, and the specified $n$ and $\beta$ parameters. The rest of the distributions are shown in Fig.~\ref{fig:Cn_dists_1}.}
    \label{fig:Cn_dists_2}
\end{figure}

In this appendix, we show the distributions of the 30 JRM observables used in the classification task. The distributions are shown in Fig.~\ref{fig:Cn_dists_1} and Fig.~\ref{fig:Cn_dists_2}.

\bibliographystyle{JHEP}
\bibliography{biblio.bib}

\end{document}